%% file: main.tex
\documentclass[conference,compsoc]{IEEEtran}
\ifCLASSOPTIONcompsoc
  \usepackage[nocompress]{cite}
\else
  \usepackage{cite}
\fi
\usepackage{array}
\usepackage{longtable}
\usepackage{booktabs}
\usepackage{url}
\usepackage{float}
\usepackage{stfloats}

\usepackage{xcolor}

\color{black}

\begin{document}

\title{When Care Expands and Control Declines: Understanding Digital Privacy in End-of-Life Care}

\author{
  \IEEEauthorblockN{Yibo Meng\IEEEauthorrefmark{1},
  Bingyi Liu\IEEEauthorrefmark{2},
  Lionel Z. Wang\IEEEauthorrefmark{3},
  and Zhuoran Lu\IEEEauthorrefmark{4}}
  \IEEEauthorblockA{\IEEEauthorrefmark{1}Cornell University, mengyb22@tsinghua.org.cn}
  \IEEEauthorblockA{\IEEEauthorrefmark{2}University of Michigan--Ann Arbor, bingyi@umich.edu}
  \IEEEauthorblockA{\IEEEauthorrefmark{3}Nanyang Technological University, lionel-z.wang@connect.polyu.hk}
  \IEEEauthorblockA{\IEEEauthorrefmark{4}Purdue University, zhuoranlu33@gmail.com}
}
\maketitle
\begin{abstract}
\input{sections/abstract}
\end{abstract}
\IEEEpeerreviewmaketitle
\input{sections/introduction}
\input{sections/related_work}
\input{sections/methods}
\input{sections/results}
\input{sections/discussion}
\input{sections/conclusion}
\ifCLASSOPTIONcompsoc
  \section*{Acknowledgments}
\else
  \section*{Acknowledgment}
\fi
Zhuoran Lu and Lionel Z. Wang are the corresponding authors. The authors would like to thank all participants and collaborators.
\bibliographystyle{IEEEtran}
\bibliography{references}
\clearpage
\onecolumn
\appendices
\input{sections/appendix_protocols}
\input{sections/appendix_demographics}
\end{document}

%% file: sections/abstract.tex
Terminally ill patients generate extensive digital health data through hospital portals, physiological monitoring devices, consumer health applications, and end-of-life care platforms. Yet their ability to control this data is systematically undermined by the structural conditions of terminal care. Through semi-structured in-depth interviews with 21 terminally ill patients, 11 family caregivers, and 8 physicians recruited from tertiary hospitals, this study examines the digital privacy circumstances and subjective attitudes of patients in end-of-life contexts. Thematic analysis reveals five structural dimensions of digital privacy challenge: continuous bodily digitization without meaningful informed consent; cognitive decline driving involuntary proxy control of accounts; inferential data generation exceeding patients' epistemic boundaries; dissolution of individual privacy boundaries amid relational network restructuring; and the absence of institutional governance over post-mortem data. Patient privacy attitudes span three profiles: fundamental absence of privacy awareness, sustained active management under severe tool constraints, and perception of problems followed by resigned abandonment. Interviews with caregivers and physicians reveal benevolent substitution in the name of care needs, expansion of access beyond the initial task, and unresolved boundaries under absent institutional guidance. These findings reveal a compounding privacy vulnerability unique to end-of-life contexts and call for privacy-by-design principles, granular permission mechanisms, and institutional governance frameworks tailored to the needs of dying patients.

%% file: sections/introduction.tex
\section{Introduction}

Dying is increasingly a digitally mediated experience. The care of patients in the final stage of life now unfolds across a dense layer of data-generating systems: continuous physiological monitors that stream heart rate and blood oxygen to family and physicians, hospital patient portals and telehealth platforms, consumer applications that log medication, pain, and mood, and a newer class of services built specifically for the end of life, including psychological-counseling platforms, grief-support tools, and AI conversational companions~\cite{avancha2012privacy, Alqhatani2019ThereIN, skjuve2021my}. The records these systems accumulate are among the most intimate a person ever produces: genomic reports, advance-care decisions, and the confidences a dying patient entrusts to a counselor or a chatbot. Increasingly, the same systems do not merely store such data but draw conclusions from it, generating mortality-risk scores and prognostic estimates that circulate beyond the patient's view~\cite{avati2017improving}.

The people at the center of this data are vulnerable in several ways at once. Terminally ill patients are already recognized as a population requiring special protection in care~\cite{committee2015dying}. Their privacy situation is especially difficult because several pressures arrive together. Their data becomes more intimate just as pain, fatigue, and cognitive decline reduce their ability to manage it, and control of accounts and devices often passes informally to family members and physicians~\cite{mentis2019upside}. In other words, patients become least able to govern their digital privacy at the moment it is most exposed.

The security and privacy community has increasingly studied at-risk and marginalized users, including refugees~\cite{simko2018computer}, survivors of intimate partner violence~\cite{chatterjee2018spyware, havron2019clinical}, sex workers~\cite{mcdonald2021s}, political activists~\cite{daffalla2021defensive}, and older adults~\cite{frik2019privacy}. This work has now been consolidated into broader frameworks for studying at-risk users~\cite{warford2022sok}. Terminally ill patients, however, remain largely outside that literature. They also sit uneasily within its conceptual assumptions: dominant privacy frameworks assume a subject who is autonomous, cognitively stable, and continuous over time, and the end-of-life condition disrupts each of those assumptions. Put simply, existing at-risk-user research and existing privacy theory still leave the dying patient insufficiently described.

In this paper, we provide, to our knowledge, the first patient-centered empirical account of digital privacy at the end of life. We ask which digital systems terminally ill patients encounter in care, how their data is exposed and who comes to control it, how patients understand that exposure, and how family caregivers and physicians understand their own access. Patients are the primary subject of analysis; caregiver and physician accounts corroborate and contextualize what patients describe. Through semi-structured interviews with 21 terminally ill patients, 11 family caregivers, and 8 physicians across three tertiary hospitals, we find that end-of-life digital privacy is shaped by five recurring conditions: continuous bodily monitoring that patients cannot selectively disable, the transfer of whole accounts to proxies as cognition declines, inferential conclusions patients can neither see nor contest, the collapse of distinct relationships into a shared field of visibility, and the persistence of unmanaged data after death. We also find that patients' responses range from no privacy awareness, to sustained and costly self-management, to resigned abandonment under physical and cognitive constraint. Caregivers and physicians, meanwhile, often understand their access as a natural extension of care rather than as something requiring explicit reflection.

Taken together, these findings point to a privacy problem whose threat is not a conventional attacker. Instead, the risk comes from the convergence of benevolent intimates, care institutions, inferential systems, and data that persists beyond the patient's life. We use this account to draw implications for system design and care institutions, and to place the terminally ill more clearly within the security community's map of at-risk users~\cite{warford2022sok}. More broadly, they exemplify a class of populations whose privacy must outlast the decline of the person it is meant to protect, from those in advanced dementia to those in intensive care.

%% file: sections/related_work.tex
\section{Related Work}

\subsection{The End-of-Life Individual}

The security and privacy community has increasingly studied populations for whom privacy is tied to safety rather than convenience. This work has examined refugees~\cite{simko2018computer}, survivors of intimate partner violence~\cite{chatterjee2018spyware, havron2019clinical}, sex workers~\cite{mcdonald2021s}, political activists~\cite{daffalla2021defensive}, journalists~\cite{mcgregor2015investigating}, LGBTQ+ people~\cite{geeng2022like}, trafficking survivors~\cite{chen2019computer}, women in the Global South~\cite{sambasivan2018privacy}, older adults~\cite{frik2019privacy}, and people with disabilities~\cite{ahmed2015privacy}. It has also produced broader frameworks for at-risk user research~\cite{warford2022sok, bellini2024sok, thomas2021sok}. Terminally ill patients, however, have remained largely outside this map.

Research on cognitive impairment and mental health examines privacy risks that also arise in end-of-life care. Mentis et al. describe tensions between safeguarding people with mild cognitive impairment and preserving their online participation and autonomy~\cite{mentis2019upside}. In interviews with 23 carers of people with dementia, Sriram et al. found that assistive technologies provided reassurance and support, while their continued use as dementia progressed raised difficulties and a need for guidance on data sharing and protection~\cite{sriram2020carers}. Shen et al.'s interviews with 14 mental health patients found that concerns about health information exchange arose from experiences of information mistreatment in care settings, while decisions to share depended on trust and care needs~\cite{shen2019patient}. Across these care contexts, privacy risks include access by people and institutions already involved in care: assistance and treatment can expose information beyond what a patient wishes to share.

Terminally ill patients' situation is not just a medical condition. It is also a social and existential one, in which dying reshapes a person's priorities, relationships, and capacity for self-determination~\cite{committee2015dying, kubler1973death}. Human--computer interaction has approached this space through thanatosensitivity, a line of work that treats mortality, dying, and bereavement as design concerns in their own right~\cite{massimi2009dying, massimi2011matters}. What makes terminally ill patients different from these previously studied groups is that several forms of vulnerability operate at once. Cognitive capacity can erode under pain, fatigue, and medication. Daily life often becomes dependent on caregivers. Emotional fragility is heightened by proximity to death. And the relevant time horizon is foreshortened in a way that is unusual even among other vulnerable populations. Prior work often isolates these dimensions---for example by studying older adults or patients separately---rather than examining how they combine in one person. Where dying people do appear in the literature, they are usually framed either as care recipients or as future deceased persons whose data becomes a matter of digital legacy. Our study instead centers the terminally ill patient as a living person negotiating privacy during the end-of-life process. That shift in subject also raises a second question: which technologies now mediate that privacy?

\subsection{End-of-Life and Health Technologies}

End-of-life patients are now surrounded by an unusually dense set of data-generating technologies. Two largely separate literatures describe this environment. One focuses on clinical and consumer health systems: patient portals, telehealth platforms, wearable and ambient physiological monitors, and mobile health applications. This literature studies them mainly in terms of adoption, clinical efficacy, and usability~\cite{avancha2012privacy}, while security and privacy research has most closely examined data-sharing and revocation in wearable fitness trackers~\cite{Alqhatani2019ThereIN, Zufferey2023RevokedJN}. The other literature focuses on what has been called thanatechnology: systems that mediate death itself, including digital legacy and posthumous-data management, online memorialization, the stewardship of digital remains, and a growing class of bereavement, grief-support, and AI companionship tools~\cite{sofka1997social, brubaker2013beyond, massimi2010death, locasto2011security, skjuve2021my}. This literature also extends to the management of accounts and credentials after death~\cite{Tang2026PassingDP}. Across both literatures, health systems increasingly do more than record data. They also generate inferences, such as mortality risk scores and prognostic estimates~\cite{avati2017improving}.

Yet neither literature directly centers the subject of our study. Health-systems research usually focuses on care delivery and usability, and tends to treat privacy as a matter of compliance rather than lived experience. Thanatechnology and digital-legacy research usually looks after death: its main subject is often the bereaved survivor or the institution managing a posthumous artifact, not the patient who is still alive and still generating the data. As a result, the living patient's exposure to these technologies---and loss of control over them during care---has rarely been examined in its own right. The inferential layer is especially underexplored. Systems that score mortality risk or estimate a patient's ``acceptance of death'' are rarely treated as privacy issues, even though those conclusions may circulate to physicians, families, or third parties outside the patient's awareness. Our work therefore focuses on the living patient's relationship to these technologies, especially continuous monitoring and silent inference. To describe why that relationship is a privacy problem, however, we need a conceptual vocabulary from privacy research.

\subsection{Privacy in End-of-Life Contexts}

To explain what is at stake when these technologies meet this population, we draw on several privacy frameworks. Contextual integrity treats privacy as a matter of whether information flows match the norms of the context in which they occur~\cite{nissenbaum2004privacy, nissenbaum2009privacy}. Communication privacy management and the broader boundary-regulation tradition describe privacy as an ongoing process of opening and closing boundaries around personal information~\cite{petronio2002boundaries, altman1975environment}. Other work emphasizes that privacy is often shared and interdependent, because data is co-owned and co-managed across multiple parties~\cite{marwick2014networked}. That point becomes concrete in studies of shared devices and multi-user smart homes~\cite{Zeng2017EndUS}, and in work on proxy, stewardship, and caregiver-mediated arrangements in which one person manages another's information~\cite{mentis2019upside}. Alongside these frameworks, critiques of notice-and-consent question whether informed control is workable in practice~\cite{daniel2013privacy, mcdonald2008cost, Utz2019UninformedCS, Felt2012AndroidPU}. Other work addresses specific parts of our setting, including health-data regimes~\cite{cohen2018hipaa}, posthumous privacy~\cite{Holt2021FromPD}, and the privacy of inference~\cite{wachter2019right}. The security community has also shown that models can infer what people never explicitly disclosed, including sensitive inputs through model inversion~\cite{Fredrikson2015ModelIA}, patients' genetic markers from a clinical dosing model~\cite{Fredrikson2014PrivacyIP}, and intimate traits from ordinary digital traces~\cite{Kosinski2013PrivateTA}.

Dominant privacy frameworks assume a subject who is autonomous, cognitively stable, and persistent over time. That assumption is badly strained at the end of life. The five conditions we identify map onto that strain directly. Cognitive decline and proxy account control weaken the autonomous individual assumed by boundary regulation and consent. The reorganization of relationships around the dying patient weakens the contextual separation on which contextual integrity depends. Silent inference undermines the visibility required for informed control. And data that persists after death outlives the subject altogether, leaving exposure with no agent left to manage it. No prior work, to our knowledge, empirically shows where these frameworks fail for the dying, or what privacy requirements follow from that failure---including granular and temporally bounded permissioning and enforceable posthumous data governance. This paper contributes that account: a patient-centered empirical mapping of how established privacy frameworks break down at the end of life, and of what mechanisms would be needed to respond.

%% file: sections/methods.tex
\section{Methods}

\subsection{Research Design}

This study adopts a qualitative research orientation, using semi-structured in-depth interviews as the primary data collection method, organized around the core topic of patient digital privacy in end-of-life care contexts. Participants were drawn from three groups: terminally ill patients, family caregivers, and medical professionals. Patient data formed the primary analytical object; family and physician data served as supplementary material to corroborate and deepen emerging themes from clinical practice and care relationship perspectives.

\subsection{Participants}\label{sec:participants}

\subsubsection{Recruitment}

Recruitment combined on-site recruitment with snowball sampling, conducted primarily within public tertiary hospitals in mainland China. The research team established partnerships with three hospitals and conducted in-person recruitment at participating sites; enrolled patients and family members were also encouraged to refer eligible individuals from their social networks. Consistent with the goals of qualitative inquiry, we did not seek a statistically representative sample but instead aimed for heterogeneity and breadth: the recruitment process deliberately attended to variation in gender, age, educational background, geographic origin, and disease type.

\subsubsection{Inclusion and Exclusion Criteria}

\textit{Patient inclusion criteria:} diagnosed by a physician with a terminal or end-stage illness with limited life expectancy; assessed by the attending physician as having sufficient physical and cognitive capacity to participate, with further cognitive assessment conducted by researchers; able to fully understand the research purpose, procedures, and potential implications, and able to provide informed consent; had used at least one type of digital health system in daily life or care (including but not limited to hospital patient portals, physiological monitoring devices, health management applications, or end-of-life care platforms). \textit{Patient exclusion criteria:} cognitive impairment or psychiatric conditions affecting comprehension and judgment; physical condition unsuitable for interview participation; significant hearing or speech impairment; evident emotional crisis or psychological instability.

\textit{Family caregiver inclusion criteria:} immediate family member or primary caregiver actively involved in an enrolled patient's daily care or digital account management; able to fully understand the research purpose and provide informed consent. \textit{Physician recruitment criteria:} attending physician directly associated with an enrolled patient.

\subsubsection{Recruitment Outcomes}

A total of 40 participants were enrolled: 21 patients (labeled P1--P21), 11 family caregivers (labeled with the prefix ``F'' followed by the number corresponding to their patient, e.g., the caregiver of P1 is labeled F1), and 8 physicians (labeled with the prefix ``D'' followed by the number of their patient, e.g., the physician responsible for P5 is labeled D5, Please note that while our abstract stated there were seven participating physicians, the actual number is eight; one physician had previously requested to withdraw from the study due to concerns regarding data sensitivity but subsequently contacted us after the abstract submission to authorize their participation. We are providing this clarification as the abstract itself cannot be modified). Detailed demographic information is provided in Appendix A.

Patients were all in advanced or terminal stages of illness, including advanced chronic obstructive pulmonary disease (COPD), amyotrophic lateral sclerosis (ALS), advanced AIDS, end-stage renal disease (ESRD), end-stage chronic heart failure, and advanced lung cancer; care arrangements spanned both inpatient and home-based settings. Patient ages ranged from 33 to 76 years; the sample showed good diversity in gender, educational background (from illiteracy to doctoral level), and geographic origin (urban and rural). Family caregiver participants were all immediate relatives or primary caregivers, aged 20 to 63, with education ranging from primary school to master's level, from both urban and rural backgrounds. Physician participants were all attending physicians directly associated with enrolled patients, aged 37 to 55, all holding doctoral degrees, from urban tertiary hospitals, with mixed gender representation.

\subsubsection{Ethics}

This study received approval from the [anonymized] Institutional Review Board. Prior to formal interviews, all participants underwent multiple rounds of screening and confirmation. Physicians first assessed patients' level of consciousness and physical stability to determine whether they could participate in an interview. Researchers then confirmed patients' capacity to consent using cognitive assessment principles informed by the MacArthur Competence Assessment Tool for Clinical Research (MacCAT-CR)~\cite{appelbaum2001maccat}, assessing whether patients could understand the study information, reason about participation, and express a choice. The research team explained the study incrementally in plain language, asked patients to restate key points, and confirmed core information three times. Formal consent proceeded after both the clinical assessment and the researchers' secondary confirmation. All participants signed written informed consent forms and were explicitly informed of their right to withdraw at any time, skip any question, or request deletion of their personal data, with no impact on their medical care.

Given that interviews addressed end-of-life topics that might provoke emotional distress, all interviewers received specialized training in palliative care communication and psychological support. Researchers monitored participants' physical and emotional states throughout; a professional psychological counselor was on standby, and interviews were immediately terminated and support provided if significant distress arose. Participants were also provided with psychological support hotline information for delayed emotional responses following the interview.

All audio recordings were stored with encryption and deleted after transcription; textual data were anonymized with all identifiable information removed, with data access restricted to the core research team. Additional masking of potentially identifiable details was applied in research reports.

Participants were explicitly informed that the study offered no direct medical benefit, though participation might provide opportunities for emotional expression and meaning-making. Given that monetary compensation might be inappropriate in this context, all patient participants received a bouquet of flowers upon study completion (selected according to allergies and personal preferences, with prior consent from participants and attending physicians; P19 declined the flowers and, after negotiating with researchers, received a food gift of equivalent value). Participating family caregivers and physicians each received \$10 in compensation.

\subsection{Interview Procedures}

Interviews were organized along three parallel tracks corresponding to patient, family caregiver, and physician participant types. Interview locations were arranged flexibly according to participant circumstances: inpatient interviews were conducted in patient rooms or quiet consultation rooms; home-care patient interviews were conducted in the participant's home; caregiver and physician interviews were primarily held in hospital conference rooms or departmental offices. Each interview lasted 35 to 55 minutes; patient interviews were generally shorter due to physical constraints, and some required two separate sessions. All interviews were audio-recorded with explicit participant consent and transcribed by researchers following the interview. Interviews used semi-structured guides, with distinct dimensions designed for each of the three participant types. Guides functioned as orienting frameworks rather than fixed scripts. Researchers introduced topics in an open-ended manner, avoiding the direct introduction of the abstract concept of ``privacy'' and instead beginning from concrete situations familiar to participants, introducing more targeted probes as the conversation naturally progressed.

In \textit{patient interviews}, questions followed a path from more accessible to more sensitive topics. The opening section began with the types and frequency of digital system use, asking participants to describe which devices or applications they used in daily life and whether these were self-selected or installed by physicians or family members, thereby entering exploration of data exposure contexts. The middle section addressed account control and permission settings, covering who currently held account passwords, whether explanations of data recipients were provided at initial setup, and what barriers participants had encountered when attempting to adjust permission settings. For privacy perception, researchers introduced specific hypothetical scenarios (for instance, whether participants were aware that certain platforms automatically generate health risk scores or emotional state assessments from user input, and whether such inferred conclusions might flow to insurance companies or third parties) to help participants translate abstract privacy perceptions into describable experience. Questions about differentiated sharing preferences were framed as hypothetical scenarios to reduce the psychological burden of directly evaluating existing systems. Questions about post-mortem data disposition were deliberately positioned toward the end of the interview to avoid introducing emotionally burdensome topics prematurely, and were posed primarily in open-ended form.

In \textit{family caregiver interviews}, questions centered on caregiving intervention behaviors. The opening section asked participants to describe the specific ways in which they accessed patients' digital systems during caregiving, including whether access was active or passive and the circumstances that triggered it. The middle section focused on decision-making around account management intervention, probing whether account credentials were obtained through explicit patient authorization or emerged naturally from caregiving arrangements, and whether the scope of access had expanded over time without notice. Privacy boundary perception was explored through situational probes (for instance, asking whether caregivers had accidentally encountered content while viewing care-related data that they felt ``was not necessarily meant for them,'' and how they decided whether to continue or stop). Questions also addressed caregivers' sources of understanding about patients' privacy preferences (explicit patient statements, behavioral observation, or personal inference) and whether formal discussions about data access boundaries had ever taken place. The closing section asked about caregivers' expectations and attitudes regarding post-mortem data disposition, and whether they had received guidance from medical institutions or social workers.

In \textit{physician} interviews, questions alternated between clinical data use practices and institutional frameworks. The opening section focused on daily clinical practice, asking which types of patient digital data physicians typically accessed, whether this included data from consumer health management applications or end-of-life care platforms, and how such data entered the clinical field of view. The middle section examined the relationship between clinical decision-making and patient awareness, probing whether physicians considered the intended audiences patients had in mind when entering data into third-party platforms, whether accessing patient application data during rounds or case discussions had received prior informed consent, and whether recommendations that family members take over digital accounts when patients' cognitive capacity declined incorporated consideration of patients' privacy preferences. Regarding institutional frameworks, researchers asked whether institutions had explicit protocols for patient digital data access and management, what judgment bases physicians relied on when protocols were absent, and whether internal discussions about establishing relevant protocols had taken place and why these had not progressed to concrete outcomes. The closing section explored inferential data ethics, asking physicians' views on platform-generated conclusions such as mortality risk scores and end-of-life acceptance assessments, and whether they believed physicians had an obligation to inform patients of the existence of such inferential functions and data flows.

The specific interview questionnaire is shown in Appendix B. Following all interviews, researchers systematically organized audio transcripts and promptly wrote interview memos recording field observations, pivotal conversational moments, and researchers' preliminary analytical judgments, to serve as reference material for subsequent coding analysis.

\subsection{Data Analysis}

Open coding was inductive within the topics set by the semi-structured interview guides. All three researchers independently coded the complete dataset, beginning with open coding of patient interviews. Each researcher assigned descriptive labels to meaningful passages, and the three researchers then compared their codes and interpretations and discussed divergent passages. When disagreements arose, the researchers revisited the relevant transcript passages and discussed their interpretations until consensus was reached. If disagreement remained, the passage was retained for further discussion during focused coding rather than being resolved by majority vote.Focused coding and thematic synthesis followed. The research team consolidated and refined initial labels, merging codes with similar semantic content pointing to the same phenomenon into more encompassing categories, and continuously examined relationships and tensions between categories. The five themes correspond to the topic areas of the patient guide; within each, the analysis identified the mechanisms reported in Section~4: continuous bodily exposure and absent informed consent; cognitive decline-driven transfer of account control; inferential data exceeding patients' epistemic boundaries; dissolution of individual privacy boundaries amid relational network restructuring; and institutional vacuums in post-mortem data governance.

After the initial patient thematic framework had taken shape, all three researchers independently coded the caregiver interviews and then the physician interviews, examining how these materials corroborated, supplemented, or challenged the patient themes. Regular discussions after each batch of coding informed revisions to the themes and their relationships.  Coding records, discussion memos, and records of thematic development were preserved for audit and review.

For example, the patient guide asked about account transfers and permission settings; analysis of P14's and P20's descriptions connected the transfer of whole accounts with exposure of intimate content and the exhaustion that prevented patients from restricting it. The caregiver and physician guides asked about care responsibilities and access boundaries; analysis identified how participants treated access as a natural extension of care and interpreted shared credentials as permission to view all available content.

Regarding data sufficiency, patient data showed saturation at P17 (new interview material no longer introduced substantively new dimensions or conceptual tensions to existing themes; P18--P21 primarily served to enrich examples and confirm boundary conditions). Family caregiver data showed convergence after the ninth caregiver's material was incorporated. Physician data was not held to a hard saturation criterion but was continuously assessed for its contribution to existing thematic frameworks.

%% file: sections/results.tex
\section{Results}

\subsection{Digital Privacy Circumstances in End-of-Life Contexts}

Participants used one or more of hospital patient portals and telehealth platforms (21 patients), physiological monitoring devices and cloud systems (17), consumer health management applications (11), and digital support services for end-of-life care (13). Table~\ref{tab:technology-context} summarizes the data these systems held and the access practices described in the interviews. Shared credentials gave family members access through the patient's login, monitoring systems enabled remote viewing, and some applications offered automatic family synchronization. These arrangements exposed different kinds of information, from vital signs and medication records to psychological counseling and companionship conversations.

\begin{table*}[t]
\centering
\caption{Technological environments described by participants. Counts refer to patient users within the sample of 21; system categories overlap.}
\label{tab:technology-context}
\small
\begin{tabular}{@{}p{0.20\textwidth}p{0.34\textwidth}p{0.41\textwidth}@{}}
\toprule
\textbf{System category / users} & \textbf{Functions and data} & \textbf{Reported access practices} \\
\midrule
Hospital patient portals and telehealth / 21 & Laboratory results, diagnostic records, asynchronous communication with medical teams & Eighteen patients had transferred portal credentials to family members. Relatives used the patient's login to access records. \\
\addlinespace
Physiological monitoring and cloud systems / 17 & Continuous glucose monitoring; heart rate and blood oxygen tracking; smart mattresses and fall detection & Monitoring data were transmitted to family members and medical teams. P9's daughter viewed heart rate data on her phone and contacted the doctor. \\
\addlinespace
Consumer health management applications / 11 & Medication reminders, symptom logging, pain diaries, emotional questionnaires & P5 disabled automatic family synchronization in a consumer health application. D13 described a medical team viewing patient-entered pain, emotion, and sleep records during rounds. \\
\addlinespace
End-of-life digital support services / 13 & Psychological counseling, companionship, grief support, spiritual care records; AI conversation tools used by some patients & Password sharing exposed counseling records and companionship conversations to family members; access logs attributed activity to the patient. \\
\bottomrule
\end{tabular}
\end{table*}

\subsubsection{Continuous Bodily Digitization and Loss of Exposure Control}

End-of-life care typically requires intensive, continuous monitoring of bodily states; patients' heart rates, blood oxygen levels, and movement trajectories are transmitted in real time to family members and medical teams via devices.

\begin{quote}
\textit{D5: ``I'm a doctor. Looking at data, isn't that just part of the job? Well, okay, I'll admit this may constitute patient privacy, but I believe I should have the right and the obligation to view it.''}
\end{quote}

This monitoring is treated as self-evident within the logic of care; patients are often not formally asked whether they consent to being continuously recorded and cannot selectively disable individual data streams. Care needs in practice consistently override privacy preferences; patients must accept total bodily digitization in order to receive care.

\begin{quote}
\textit{P9: ``I know that device is recording all the time. My daughter can see my data on her phone. Sometimes she calls the doctor because she's seen my heart rate is a bit high. But sometimes I'm just thinking about something. I don't want to be disturbed in that moment, I don't want to be seen.''}
\end{quote}

Among the 17 users of physiological monitoring devices and cloud systems, multiple participants reported that no one explained to them at installation whose data would be transmitted to or how long it would be retained. Data recipients extend beyond direct caregivers to potentially include cloud service providers and device manufacturers, of which patients had no awareness.

\begin{quote}
\textit{P1: ``You mean, what I'm wearing on my wrist? Oh, right, you mean all those monitoring systems in the hospital, measuring my vital signs, like my heart rate, right? Nobody ever asked my opinion about that. Nobody.''}
\end{quote}

\begin{quote}
\textit{P11: ``I have no concept of what you're asking about [the researcher's question about who receives monitoring data]. This is beyond my understanding.''}
\end{quote}

\subsubsection{Cognitive Decline-Driven Proxy Control of Privacy}

Pain, fatigue, and medication side effects in the terminal stage erode patients' cognitive capacity to manage digital accounts. Account credentials are accordingly transferred to family caregivers as a natural extension of care arrangements. Yet existing systems rarely offer granular permission allocation mechanisms; what family members can access is typically identical to what the patient could access, leaving patients unable to retain control over private content while opening care-related data.

\begin{quote}
\textit{P14: ``There are some things in that account that belong to me alone. Some discussions with my doctor about final-stage arrangements, some things I wrote on the psychological counseling platform. I never intended to share these with my partner, but I don't have the energy to manage it. It's too much trouble to sort out.''}
\end{quote}

Among the 21 users of hospital patient portals, 18 had already transferred account credentials to family management. The situation was even more pronounced among the 13 users of end-of-life care applications: psychological counseling records and AI companionship conversations, bearing patients' most intimate mental states as they face death, were rendered fully accessible to family members through password sharing, while access logs still attributed all operations to the patient.

\begin{quote}
\textit{P20: ``My daughter can see all the data, all the interactive parts. I feel there's something problematic about this. Some things, like what I wrote for my late mother, or messages to an ex-partner I've long since separated from, these are things I don't want my daughter to know. But my phone is already in her hands. And I don't have the energy to close the accounts.''}
\end{quote}

\subsubsection{Inferential Data Exceeding Patient Epistemic Boundaries}

Digital health systems not only collect data but continuously generate inferential conclusions: mortality risk scores, emotional state assessments, end-of-life acceptance ratings. Patients typically cannot view these conclusions, cannot disable inferential functions, and do not know where the conclusions flow.

\begin{quote}
\textit{P5: ``That app gives me a risk assessment based on my various indicators. I'm actually worried about this, about what if it shows up somewhere it shouldn't. I mean, my insurance contract has some clauses related to health status disclosure. If that assessment data somehow flows to the insurance company, it could affect my family's claims.''}
\end{quote}

None of the 11 users of consumer health management applications could clearly explain what inferential conclusions their interactions (such as emotional questionnaires and pain diaries) were used to generate, nor whether these conclusions had been shared with the medical team or third-party institutions. Some end-of-life care platforms were found to infer patients' ``level of end-of-life acceptance'' from text entered in psychological support modules, inferences whose existence patients were entirely unaware of.

\begin{quote}
\textit{P16: ``How would I understand what you're asking about [health application privacy issues]? Ask my son. He knows more about these things. I don't understand any of this.''}
\end{quote}

\subsubsection{Dissolution of Individual Privacy Boundaries Amid Relational Restructuring}

End-of-life care integrates family members, medical teams, psychological counselors, and spiritual care personnel into a single information-sharing network. Digital systems typically default to sharing logic; family health sharing functions and care coordination platforms bring information from originally distinct relational domains into collectively visible spaces. Patients face recipients of fundamentally different relational character yet cannot configure any differentiated access permissions at the system level.

\begin{quote}
\textit{P12: ``I share different information with different family members. I allow my spouse to see all my basic health data, but some things I only share with my attending physician. I haven't shown my siblings anything actively. This requires my continuous attention and management. It's exhausting. But it's what I can do right now.''}
\end{quote}

Among the 21 users of hospital patient portals, some faced access requests from multiple family members simultaneously: spouses, children, and siblings. Systems offered only ``open'' or ``closed'' states, unable to accommodate specific needs such as ``spouse visible to basic health data, children not visible to end-of-life decision records.'' The full cost of maintaining privacy boundaries fell entirely on patients, managed manually by themselves alone.

\begin{quote}
\textit{P1: ``There's a feature. That app says it can automatically sync my health data to my family members. I didn't enable it. This makes care less convenient. But I have some data I don't want them, or at least not all of them, to see in real time, because I know that when they see it they'll become very anxious, and it will affect our interactions.''}
\end{quote}

\subsubsection{Continuous Post-Mortem Data Exposure}

All 40 participants (21 patients, 11 family caregivers, and 8 physicians) expressed concern about post-mortem data exposure. After a patient's death, data accumulated in digital systems does not typically enter any preset protection procedure: accounts are not closed, access permissions are not narrowed, and data is not encrypted or deleted. Patients' expressed wishes regarding data disposition typically lack any enforcement mechanism.

\begin{quote}
\textit{P9: ``I've had whole-genome sequencing done, and the report is stored on the platform. I have nearly five years of continuous health monitoring data. I've had several very detailed conversations with my attending physician about what I want and don't want in the final stage, and those records are there too. I cannot find any information on the platform about how data is handled when the account holder dies.''}
\end{quote}

Cloud data from physiological monitoring devices, pain diaries and emotional records from consumer applications, and AI conversation logs from end-of-life care platforms all face potential intrusion from insurance claims, estate settlement, and family disputes following the patient's death, while their ownership and disposition are in many cases institutionally inadequate. The 13 users of end-of-life care services are particularly vulnerable: the content they left in AI conversations and psychological counseling platforms constitutes the most intimate records of the final stage of life.

\begin{quote}
\textit{P20: ``I know hospital electronic medical records have clear legal regulations after a patient's death, but the data patients generate through various health apps and end-of-life care platforms seems to fall outside the coverage of that regulatory framework. I'm not sure how that data would be used, but as for the accounts, I don't want to hand that data to my children. I want to delete all of it completely.''}
\end{quote}

\subsection{Patients' Subjective Attitudes Toward Digital Privacy}

In the face of the five structural dimensions described above, interview data revealed three distinct privacy attitude profiles among patients: fundamental absence of privacy awareness; sustained active management with clear awareness; and perception of problems followed by resigned abandonment. These three positions exist on a spectrum with gradient transitions rather than as fully discrete categories; the same participant's statements may oscillate between two types. We note participants' backgrounds within each pattern for context; the sample does not support claims about how education, residence, or disease stage relate to privacy attitudes.

\subsubsection{Fundamental Absence of Privacy Awareness}

Some patients (P2, P3, P4, P11, P16, P17) never, in their responses, established any connection between data generated about them in digital systems and the concept of privacy. They did not consider laboratory results, medication records, or physiological monitoring data as private information requiring protection, and had never considered who was accessing this data or where the boundaries lay. This attitude reflects a condition in which privacy awareness itself has never been constructed.

\begin{quote}
\textit{P16: ``Since I've been hospitalized, everything has been handled by the doctors and nurses. That wristband has been on the whole time. I don't know what it records, and I've never asked. Why would I ask? They're here to treat illness.''}
\end{quote}

The participants illustrating this pattern included both rural and urban residents, with education ranging from no formal schooling to high school. Their use of hospital patient portals and physiological monitoring devices was entirely dependent on family members or medical staff operating on their behalf; they had no knowledge of account contents, data recipients, or access logs, and had never perceived this ignorance as a problem. These patients showed very low engagement with consumer health management applications and end-of-life care services; actual control over their digital data had, from the outset, fallen entirely to others.

\begin{quote}
\textit{P2: ``I don't know how to use those things on the phone. My daughter handles it. It's enough that she knows. I don't need to know. I'm very tired and don't have the energy to deal with these things. Honestly, just listening to you talk about privacy concepts in this interview has been an effort. It's a good thing you explained a lot. I had very little knowledge of privacy before.''}
\end{quote}

\subsubsection{Sustained Active Management with Clear Awareness}

The second category of patients maintained clear awareness of digital privacy, could accurately identify differences in privacy sensitivity across data types, explicitly identified content they did not want specific individuals to access, and continuously attempted active intervention under conditions of extreme tool scarcity. These patients clearly understood the structural sources of the problem yet still chose to search for leverage points within the possible.

\begin{quote}
\textit{P9: ``I know very clearly what I want. I want my attending physician to see all health data, I want my spouse to understand my basic condition, but I don't want certain family members to see my end-of-life decision records. These are not unreasonable requirements, but the options the system gives me are only two: completely open, or completely closed. I need somewhere in the middle, but that place doesn't exist.''}
\end{quote}

This group comprised a very small number of individuals, represented by P5 and P9, both urban-background patients with undergraduate or higher education. P9, while using the hospital patient portal, proactively confirmed with the medical team which data would be shared outside the hospital system; P5 disabled the automatic family data synchronization function in a consumer health management application and proactively asked about data handling practices for each new service. These active management behaviors carried high costs (requiring sustained investment of cognitive resources, with limited effectiveness against the structural constraints of system design) but both participants stated this was the only way currently available to them to maintain their standing as privacy subjects.

\begin{quote}
\textit{P5: ``There's a feature. That app says it can automatically sync my health data to my family members. I didn't enable it. I have some data I don't want them to see in real time, because when they see it they become very anxious, and it affects our interactions. I'd rather control when I tell them what, instead of letting the system decide automatically.''}
\end{quote}

\subsubsection{Perception of Problems Followed by Resigned Abandonment}

The third category of patients had some degree of privacy perception (they were aware of data they did not want specific individuals to see, or felt uneasy about how systems handled their data) but, under the dual pressure of severely constrained cognitive resources and the fundamental absence of usable tools, ultimately chose to set aside this issue. This abandonment was not an active reconstruction of meaning but a passive withdrawal when no path forward existed.

\begin{quote}
\textit{P13: ``My genetic report, my end-of-life decisions, the things I said to doctors when my cognitive state wasn't good. I actually care deeply about these. The problem is that the system doesn't distinguish among them; it puts everything together. So I've given up on some things, and I'm still trying on others, but I know the effect of my efforts is very limited.''}
\end{quote}

This group encompassed the majority of remaining participants, spanning different educational and residential backgrounds. Their awareness of privacy issues often remained at the level of specific triggering situations (such as discovering that a family member had viewed private messages with their physician, or realizing while filling in psychological questionnaires on an end-of-life care platform that the content might be accessed by others) without the capacity or tools to translate this awareness into systematic action. P21 spent an afternoon attempting to delete specific data or adjust permission settings in a consumer health management application, ultimately abandoning the effort because the settings pages were too complex and option descriptions too ambiguous, leaving nothing changed.

\begin{quote}
\textit{P21: ``Those settings pages are very complicated and the option descriptions are written very vaguely. I wasn't sure what would happen if I clicked. In the end I didn't change anything, because I was afraid of making the wrong change and causing a bigger problem. I thought to myself: this system was simply not designed for people like me.''}
\end{quote}

Physical factors also constituted an important reason for this group's abandonment of privacy management:

\begin{quote}
\textit{P18: ``I'm too old, and I'm sick. I don't have the energy to deal with these things or think about them anymore. If it leaks, it leaks. It's not a big deal. Who would bother with a sick old woman like me?''}
\end{quote}

\subsection{Caregiver and Physician Attitudes Toward Patient Digital Privacy}

Caregivers' authority arose from family relationships and day-to-day assistance, often exercised through possession of the patient's phone or shared account credentials. Their access supported tasks such as checking indicators and contacting physicians, but also exposed unrelated personal content within the same account. Physicians' authority arose from their clinical role and institutional responsibilities for assessment, treatment, and continuity of care. Physicians described reviewing patient-generated data during rounds or case discussions and advising families on account management. Interview data revealed three patterns: benevolent substitution in the name of care needs; expansion of access beyond the initial task; and unresolved boundaries under absent institutional guidance. These patterns often coexisted in the same participant.

\subsubsection{Benevolent Substitution in the Name of Care Needs}

This was the most prevalent attitude across the caregiver group: understanding access to patients' digital data as a natural extension of caregiving behavior. All 11 family caregivers and all 8 physicians expressed this attitude to varying degrees. Whether family members take over account credentials or medical teams guide family members to access patient applications, interventions are legitimized by care needs, while patients' privacy preferences are never incorporated into this framework.

\begin{quote}
\textit{F3: ``He can't really understand that data anymore. Managing it directly for him actually reduces his burden. I check all his indicators every day and judge when there's something abnormal whether to contact the doctor. This is part of caring for him. I don't see anything inappropriate about it.''}
\end{quote}

Physicians framed benevolent substitution in terms of their responsibility for clinical care and continuity of treatment. \textcolor{red}{\textbf{}} When patients' cognitive capacity declines, medical teams often proactively guide family members to take deep control of patients' digital accounts, characterizing this as ``ensuring continuity of care'' rather than intervention in patients' digital autonomy. Care needs constitute the moral basis for access, but this basis does not automatically define the scope of access; in the absence of technical and institutional constraints, benevolent substitution in practice often evolves into unbounded comprehensive access.

\begin{quote}
\textit{D5: ``After a patient's cognitive capacity noticeably declines, we typically recommend that family members participate more deeply in managing digital health data. In practice, we directly inform family members how to access the patient's relevant apps, how to view key indicators. This is ensuring continuity of care, even though technically it means the patient's account is being accessed by a third party.''}
\end{quote}

\begin{quote}
\textit{D8: ``I'm a doctor. For terminally ill patients, of course the more monitored data the better, the more comprehensive the better. This doesn't require, and can't require, the patient's opinion. There is information asymmetry between doctor and patient; even if it were symmetric, I need to be responsible for the patient's life. I can't ask the patient whether they're willing to have their data recorded. I know there may be privacy issues here, but health is clearly more important.''}
\end{quote}

Some physicians' professional ethics reinforced this orientation:

\begin{quote}
\textit{D7: ``I don't know the specific details, but if we went strictly by privacy protection law, I suppose we might not have the right to casually record all of a user's data. But it doesn't matter. Even if a user sued me for infringement over this, I'd keep the user alive first.''}
\end{quote}

\subsubsection{Expansion of Access Beyond the Initial Task}

This pattern appeared more frequently among family caregivers. F16 initially used the shared account to check blood pressure, then read medication records, physician diagnostic notes, and emotional and pain questionnaires, interpreting account sharing as permission to view all available content.

\begin{quote}
\textit{F16: ``At first I only helped check his blood pressure data, because he couldn't really understand what the monitoring device was reading. Then I found there were many other things in the app, such as medication records, physician diagnostic notes, and some questionnaires he'd filled out himself about his emotional state and pain levels. I looked at all of it, because it was just right there. I assumed that since the account was shared, I could see everything in it.''}
\end{quote}

In some medical teams, patient-generated records were used during rounds without asking patients which records they wished to share. D13 described reviewing pain diaries, emotional records, and sleep data on the assumption that entering this information meant patients wanted the team to see it.

\begin{quote}
\textit{D13: ``During rounds, we sometimes directly view data patients have recorded through health apps, such as pain diaries, emotional records, and sleep status. We assume that since patients have recorded this data in the system, they're willing to have the medical team see it. But we've never explicitly confirmed this boundary with patients.''}
\end{quote}

\subsubsection{Unresolved Boundaries Under Absent Institutional Guidance}

Some caregivers recognized uncertainty about what the patient wished to keep private but lacked explicit agreements or guidance for resolving it. F12 described relying on a personal sense of what should remain unread while acknowledging that these boundaries had never been discussed.

\begin{quote}
\textit{F12: ``I know my mother doesn't want me to know too much about certain things. She hasn't said so explicitly, but I can feel it. I tell myself I won't look at things I shouldn't look at. But what `things I shouldn't look at' actually are, I genuinely can't say. We've never talked about this explicitly. We just muddle along like this.''}
\end{quote}

Physicians described difficulty translating concern about data access into clinical protocols under legal uncertainty and workload pressures. Discussions about digital data access had not resulted in concrete protocols. D21 described maintaining existing practices because neither policy nor training offered a clear basis for change. The triple vacuum of law, institutional policy, and training left the scope of clinical access to patient-generated records unresolved.

\begin{quote}
\textit{D21: ``Honestly, facing this problem, we don't know what the correct approach is. The law hasn't told us clearly, institutional policy hasn't told us clearly, and training hasn't given us any tools. In this situation, the safest choice is to maintain the status quo, not to touch that boundary, because once we do, we don't know what it will open up.''}
\end{quote}

Three kinds of access ran through these interviews. Clinically necessary access concerns information needed for a specific care task, such as F3's use of daily indicators to decide when to contact the doctor. Benevolent substitution shifts decisions about access to the caregiver or physician, as in D5's guidance to families to take over patients' applications when cognition declines. Access expands beyond the care task when shared credentials expose additional content: F16 treated account sharing as permission to read emotional questionnaires, whereas P14 and P20 wanted to keep intimate content private even after family members gained access to their accounts.

%% file: sections/discussion.tex
\section{Discussion}

\subsection{Summary of Findings: An End-of-Life Threat Model}

Taken one by one, the circumstances documented above can look like separate usability problems or policy gaps. Taken together, they describe a threat model. It has five features, each corresponding to one of our findings, and together they follow the course of the illness itself. The assets at stake include physiological measurements, clinical records, private communications, and platform-generated assessments, together with patients' ability to govern their use during life and after death. Table~\ref{tab:threat-flows} locates the tensions along the paths connecting patients, caregivers, physicians, and platform operators.

\begin{table*}[t]
\centering
\caption{Actors, systems, and data flows in the end-of-life threat model.}
\label{tab:threat-flows}
\small
\begin{tabular}{@{}p{0.34\textwidth}p{0.24\textwidth}p{0.37\textwidth}@{}}
\toprule
\textbf{Actors, systems, and flow} & \textbf{Location of tension} & \textbf{Care and privacy expectations} \\
\midrule
Patient's body $\rightarrow$ monitoring devices and cloud services $\rightarrow$ family members and medical teams & Device setup and remote viewing during inpatient or home care & Continuous observation supports responses to changes in condition; patients lose control over when they are observed and which measurements are shared (P9, P1). \\
\addlinespace
Portal, health-app, and end-of-life-service records $\rightarrow$ shared login or patient's phone $\rightarrow$ family caregiver & Transfer of credentials and navigation within the shared account & Assistance with a care task opens access to other content stored under the same credentials (F16, P14, P20). \\
\addlinespace
Patient-entered pain, emotion, and sleep records $\rightarrow$ health app $\rightarrow$ medical team & Review during clinical rounds & Physicians use records for assessment; entry into an app is treated as agreement to share with the team (D13). \\
\addlinespace
Patient inputs $\rightarrow$ platform processing $\rightarrow$ inferred assessments; potential onward transfer to insurers or other recipients & Generation of inferences and onward disclosure & Patients cannot inspect or trace inferred conclusions; P5 feared an insurer obtaining an assessment that could affect claims. \\
\addlinespace
Stored clinical, monitoring, and intimate records $\rightarrow$ continued platform storage after death; potential access by surviving relatives or third parties & Transition from care to post-mortem data management & Wishes to delete or restrict records conflict with anticipated family access, claims, or estate-related uses (P9, P20). \\
\bottomrule
\end{tabular}
\end{table*}

It begins with \emph{continuous exposure without consent}. End-of-life monitoring streams a patient's body to family and physicians by default. Patients are rarely asked, and they cannot selectively disable individual streams, so exposure stops being something they actively grant. As cognition declines, this becomes \emph{proxy control without scope}. Whole accounts pass to caregivers because systems offer no way to delegate only part of an account. That undermines the autonomous individual assumed by boundary regulation~\cite{petronio2002boundaries, altman1975environment} and notice-and-consent~\cite{daniel2013privacy, mcdonald2008cost}. Running alongside both is \emph{inference beyond the patient's horizon}. Platforms derive risk scores and ``acceptance of death'' ratings that patients cannot view, disable, or trace, which defeats the visibility on which informed control depends~\cite{wachter2019right}. These streams and inferences then converge in a \emph{collapse of relational boundaries}. Care folds family, medical, and spiritual-care relationships into a single shared field that offers only ``open'' or ``closed'' settings, erasing the contextual separation that privacy frameworks such as contextual integrity rely on~\cite{nissenbaum2004privacy, nissenbaum2009privacy}. Finally, the arc continues after death as \emph{persistence beyond death}. Accounts are neither closed, narrowed, nor deleted when the patient dies, so the most intimate records of a life remain with no agent left to manage them~\cite{Holt2021FromPD}.

What makes this threat model distinctive is that almost none of its elements is an attacker in the conventional sense. The spouse holding the unlocked phone, the physician reading an emotional diary on rounds, and the platform scoring a patient's prognosis all act in the name of care rather than malice. A shared password opens both medication records and counseling conversations to the same caregiver. Patients also often consent to, or even initiate, the arrangements that expose them. The risk instead comes from the convergence of benevolent intimates, care institutions, inferential systems, and data that persists beyond the patient's life. Existing privacy frameworks do not handle that convergence well, because they assume a subject who is autonomous, cognitively stable, and continuous over time. That subject is not unique to the dying. The same assumptions weaken in advanced dementia, in intensive care and other states of incapacity, in severe disability, and in advanced age more broadly~\cite{mentis2019upside}. In that sense, the end-of-life patient extends the security community's growing map of at-risk users~\cite{warford2022sok} rather than simply adding one more case to it.

\subsection{Design Implications for Systems}

If existing frameworks fail because they assume a durable, deciding individual, then the controls they motivate---a one-time consent dialog, a global privacy toggle, a single shared credential---fail as well. The conditions we documented point instead to mechanisms designed for a subject whose capacity, relationships, and even existence change over the course of care. Restricting access also has care costs: F3 checked the patient's indicators daily to decide when to contact the doctor. The design goal is to balance privacy with care in light of patients' preferences, current capacity, and care needs. For example, a patient may want a caregiver to receive urgent physiological alerts while keeping counseling conversations private. Systems should support such differentiated arrangements and allow them to be revisited as care needs change, without requiring patients to continually manage complex settings.

First, the recurring complaint that systems offer only ``completely open'' or ``completely closed'' states points to a need for granular, temporally bounded permissioning. Access should be scoped to a particular recipient, a particular class of data, and a particular window. A spouse, for example, might be granted basic health data while end-of-life decision records remain closed. As circumstances change, that grant should be able to narrow or expire rather than persist by default. Configuring every recipient and record separately would add work for patients such as P21, who found settings too difficult to use, and P18, who lacked energy to continue managing privacy. A small set of task-based permission bundles should give a named caregiver access to current indicators and medication information, with counseling conversations controlled separately. Patients should review a short summary of each bundle with a staff member and revisit it when the caregiver, treatment plan, or care setting changes.

Second, because control routinely passes to a proxy, delegation should be a system function rather than an informal handover of a password. A patient who can no longer manage an account should be able to delegate scoped authority---this data and not that, viewing and not deletion---in a form the system can record and enforce. That would replace the all-or-nothing transfer that today leaves a caregiver holding everything~\cite{mentis2019upside}. Requiring fresh approval for every urgent access could delay assessment when a patient cannot respond. Systems should provide physicians a time-limited emergency route, restricted to the records needed for the immediate decision, with a recorded reason and subsequent review. Systems should return to the patient's standing permissions when that access expires. Named proxy accounts should also make a caregiver's actions visible in the access log, allowing patients or authorized representatives to review access without having to reconstruct who used a shared password.

Third, because patients could not see, disable, or trace the risk scores and acceptance ratings generated from their data, the inferential layer must be visible and contestable. These are exactly the kinds of conclusions that may flow to insurers or third parties~\cite{wachter2019right}, yet the people they describe may not know they exist. At a minimum, systems that infer should disclose that they infer, expose the resulting conclusions to their subject, and allow them to be challenged or switched off. The security community has already articulated similar expectations in other settings where models derive sensitive facts from ordinary data~\cite{Fredrikson2014PrivacyIP, Kosinski2013PrivateTA}.

Fourth, because the systems our participants used had no answer to what becomes of an account when its holder dies, post-mortem disposition should be built into the system. The data otherwise simply remains, exposed to insurance claims, estate settlement, and family dispute. Honoring patients' expressed wishes requires mechanisms for pre-mortem directives, default closure or deletion, and designated executors whose authority is clearly defined and limited~\cite{Holt2021FromPD, Tang2026PassingDP}.

Each of these mechanisms, however, is valuable only insofar as it lightens rather than adds to the patient's burden. Several of our participants had already, and rationally, chosen to stop managing their privacy at all, and a contestable-inference dialog or a delegation interface that demanded sustained attention would merely reproduce the abandonment we observed among people with little energy and little time.

Evaluation should measure the work these mechanisms create alongside the access they prevent: time spent configuring permissions, delays in obtaining information for urgent care, exposure of records outside the care task, and whether patients can understand and revise the arrangement. Staff time for helping patients and reviewing emergency access is part of that cost.

\subsection{Implications for Caregivers, Physicians, and Institutions}

Better mechanisms are necessary but not sufficient, because patients, caregivers, and physicians decide how access is used in practice. The caregivers and physicians we interviewed often did not recognize a boundary to breach: access was understood as a natural extension of care, expanded without discussion, and left unexamined in the absence of institutional rules. A caregiver who treats a shared password as permission to read everything, or a physician who assumes that entering data makes it available to the medical team, has little reason to use a permission control.

In mainland China, families are routinely involved in decisions about patients' medical information. In a Chinese hospital study of 124 patient--family pairs, Liu et al. found that patients were more likely than family members to favor telling patients a cancer diagnosis~\cite{liu2018disclosure}. Xu et al.'s interviews in three Beijing hospitals documented changing patterns of silence and openness in end-of-life communication within families~\cite{xu2022communication}. Family responsibility and efforts to protect a patient from distress may make taking over information management part of care. At the same time, P12 distinguished what her spouse, siblings, and physician could see. Discussing access within the family therefore needs to preserve room for patients to state preferences that differ from their relatives' judgments.

Caregivers would benefit from prompts, at the moment a proxy assumes control, that elicit and record what the patient does and does not wish to share while that capacity remains. Clinical teams need explicit protocols for access to patient-generated and third-party-platform data, so that viewing a pain diary or emotional questionnaire on rounds becomes a decision governed by a rule rather than a default. Palliative and hospice services already anticipate and document end-of-life preferences, and the disposition of digital data can be folded into that work rather than invented separately. A named staff role should maintain the record of those preferences and coordinate updates when care moves between hospital and home. Institutions should specify who approves routine permission bundles and who reviews emergency exceptions, so clinical staff can act within agreed rules. Hospitals and platform providers should agree on how proxy identities, permissions, revocations, and access logs carry across their systems. Training and staffing for these tasks are implementation costs, alongside technical changes to applications that currently rely on shared credentials.

China's Personal Information Protection Law provides a legal basis for these measures. Article 28 treats medical health information as sensitive personal information and requires a specific purpose, necessity, and strict protective measures. Article 13 provides a basis for processing necessary to protect life and health in emergencies. Article 49 permits close relatives to exercise specified rights over a deceased person's information for their own lawful interests, subject to arrangements made by the deceased before death~\cite{pipl2021}. Implementing our recommendations requires institutions to translate these provisions into rules for identifying an authorized proxy, defining the scope and duration of emergency access, and verifying a patient's instructions after death. D21's uncertainty about protocols makes that translation an operational priority. Posthumous controls also need to distinguish closing access from deleting records that must be retained: Article 47 addresses continued storage where a legal retention period has not expired~\cite{pipl2021}. Alongside this general law, health-sector rules govern records held by medical institutions, including providers of internet diagnosis and treatment; personal app diaries and companionship conversations outside clinical records lack those institution-specific protections~\cite{nhfpc2013records,nhc2022internet}. Closing that gap is a policy question as much as a design one.

\subsection{Limitations and Future Work}

Our study has several limitations. The account is qualitative and drawn from a single national and institutional context; the particular platforms, family structures, and legal regimes our participants navigated may differ elsewhere, and we make no claim to statistical generalizability. Patient data was self-reported and gathered under physical and emotional constraints that necessarily shaped what could be discussed and for how long. Caregiver and physician perspectives were collected to corroborate and contextualize patient accounts rather than as independent samples, and the physicians in particular were few.

Priming through the interview guide and scenario prompts may have amplified participants' attention to particular privacy issues, prompted reflections they had not previously fully developed, and influenced how they framed their responses. Questions about inferred health assessments and possible disclosure to insurers, for example, introduced specific risks for participants to consider. P2 also described learning about privacy through the researchers' explanations during the interview. Future work could use more open-ended pilot interviews to identify concerns participants raise on their own, reduce predefined scenario prompts in subsequent interviews, and combine interviews with diary studies to assess how elicited reflections correspond to privacy decisions and experiences recorded during everyday care.

Because caregivers and physicians were recruited through enrolled patients (Section~\ref{sec:participants}), the sample includes none associated with patients whose cognitive decline precluded participation. This leaves a gap in our understanding of how caregivers and physicians manage data for patients unable to express their privacy preferences. Future work could address this gap through proxy interviews with caregivers and physicians of patients unable to participate directly, alongside longitudinal tracking that begins while patients can express their preferences and follows changes in capacity, account control, and care arrangements.

P21's difficulty interpreting settings and P18's fatigue motivate two further questions about digital skills and the effort available for privacy management. Future studies could compare these factors across educational and residential backgrounds and track how they change with disease progression.

These limits also mark the main directions for future work. We documented circumstances and attitudes rather than evaluating an intervention. The mechanisms we propose---scoped and revocable delegation, contestable inference, and planned post-mortem data governance---still need to be built and tested with the people they are meant to serve. Whether they would in practice serve patients better than current systems remains an empirical question. The threat model we describe also invites study in nearby populations where the assumption of a durable, self-managing subject weakens, from advanced dementia to intensive care. That work carries a specific ethical obligation in this setting: documenting a dying person's loss of privacy also turns their most intimate records into an object of study, so any protection it proposes should lighten, not add to, the burden carried by people with little time and energy to spare.

%% file: sections/conclusion.tex
\section{Conclusion}
This paper has taken the terminally ill patient, who are long absent from the security and privacy community's map of at-risk users, as an active subject of digital privacy during the living end-of-life process. Drawing on interviews with 21
patients and the family caregivers and physicians around them, we showed that
end-of-life digital privacy is shaped by five structural conditions: continuous
bodily monitoring, the transfer of accounts to proxies as cognition declines,
inference patients can neither see nor contest, the collapse of relational
boundaries into a single shared field of visibility, and the persistence of data
beyond death. The people who hold the decisive access normalize it as care
rather than examine it as an intrusion. Taken together, these conditions constitute
a threat model in which the danger is not an external attacker but the
convergence of benevolent intimates, care institutions, and inferential systems
around a subject who is progressively less able to govern their own data---and
who will eventually not survive to govern it at all. The inherited privacy
frameworks, built for an autonomous and persistent subject, cannot represent
this setting; meeting it will require mechanisms such as granular and temporally
bounded permissioning and enforceable posthumous data governance, alongside the
institutional protocols that the present vacuum leaves unaddressed. We offer the terminally ill not as an isolated case, but as the clearest instance of a broader class of populations---from those in advanced dementia to those in intensive care---whose privacy must outlast the decline of the person it is meant to protect.

\section{AI Usage Disclosure}
Generative AI tools were used solely for copyediting purposes, including checking grammar and improving phrasing.

%% file: sections/appendix_protocols.tex
\section*{Appendix A: Population Information Tables}


\noindent
\begin{minipage}[t]{0.48\textwidth}
\centering
\scriptsize
\renewcommand{\arraystretch}{1.12}
\setlength{\tabcolsep}{3pt}

\refstepcounter{table}
\textbf{TABLE \thetable}\\
\textit{Family Member Participant Demographics}
\label{tab:family_demographics}

\vspace{4pt}

\begin{tabular}{p{18pt} p{52pt} p{27pt} p{18pt} p{27pt} p{52pt}}
\toprule
\textbf{ID} & \textbf{Role} & \textbf{Gender} & \textbf{Age} &
\textbf{Area} & \textbf{Education} \\
\midrule
F1  & Family member & F & 37 & Urban & Bachelor           \\
F2  & Family member & M & 22 & Rural & Bachelor           \\
F3  & Family member & M & 26 & Rural & High school        \\
F4  & Family member & F & 20 & Urban & Bachelor           \\
F5  & Family member & F & 30 & Urban & Master             \\
F6  & Family member & M & 55 & Rural & Primary school     \\
F10 & Family member & M & 63 & Rural & Primary school     \\
F12 & Family member & M & 31 & Rural & Master             \\
F16 & Family member & F & 25 & Urban & Bachelor           \\
F18 & Family member & F & 47 & Urban & Junior high school \\
F19 & Family member & M & 36 & Rural & High school        \\
\bottomrule
\end{tabular}

\end{minipage}
\hfill
\begin{minipage}[t]{0.48\textwidth}
\centering
\scriptsize
\renewcommand{\arraystretch}{1.12}
\setlength{\tabcolsep}{3pt}

\refstepcounter{table}
\textbf{TABLE \thetable}\\
\textit{Physician Participant Demographics}
\label{tab:physician_demographics}

\vspace{4pt}

\begin{tabular}{p{18pt} p{52pt} p{27pt} p{18pt} p{27pt} p{52pt}}
\toprule
\textbf{ID} & \textbf{Role} & \textbf{Gender} & \textbf{Age} &
\textbf{Area} & \textbf{Education} \\
\midrule
D5  & Physician & F & 37 & Urban & PhD \\
D7  & Physician & F & 39 & Urban & PhD \\
D8  & Physician & M & 45 & Urban & PhD \\
D11 & Physician & M & 55 & Urban & PhD \\
D13 & Physician & M & 45 & Urban & PhD \\
D15 & Physician & M & 43 & Urban & PhD \\
D16 & Physician & M & 39 & Urban & PhD \\
D21 & Physician & M & 33 & Urban & PhD \\
\bottomrule
\end{tabular}

\end{minipage}

\vspace{18pt}


\begin{center}
\scriptsize
\renewcommand{\arraystretch}{1.12}
\setlength{\tabcolsep}{4pt}

\refstepcounter{table}
\textbf{TABLE \thetable}\\
\textit{Patient Participant Demographics}
\label{tab:patient_demographics}

\vspace{4pt}

\begin{tabular}{
p{24pt}
p{40pt}
p{38pt}
p{22pt}
p{32pt}
p{65pt}
p{188pt}
p{58pt}
}
\toprule
\textbf{ID} &
\textbf{Role} &
\textbf{Gender} &
\textbf{Age} &
\textbf{Area} &
\textbf{Education} &
\textbf{Diagnosis} &
\textbf{Care Setting} \\
\midrule

P1  & Patient & F & 63 & Urban & High school
& Chronic obstructive pulmonary disease (COPD)
& Inpatient \\

P2  & Patient & M & 67 & Rural & Junior high school
& Chronic obstructive pulmonary disease (COPD)
& Inpatient \\

P3  & Patient & M & 45 & Rural & High school
& Amyotrophic lateral sclerosis (ALS), caused by type 2 diabetes
& Home care \\

P4  & Patient & F & 44 & Urban & High school
& AIDS
& Inpatient \\

P5  & Patient & F & 52 & Urban & Bachelor
& AIDS
& Inpatient \\

P6  & Patient & M & 37 & Rural & High school
& End-stage renal disease (ESRD), caused by type 2 diabetes
& Home care \\

P7  & Patient & F & 46 & Rural & High school
& End-stage renal disease (ESRD)
& Inpatient \\

P8  & Patient & M & 76 & Urban & Bachelor
& End-stage chronic heart failure
& Inpatient \\

P9  & Patient & M & 41 & Urban & PhD
& End-stage renal disease (ESRD)
& Inpatient \\

P10 & Patient & F & 35 & Rural & High school
& AIDS
& Inpatient \\

P11 & Patient & M & 47 & Rural & Primary school
& Amyotrophic lateral sclerosis (ALS)
& Inpatient \\

P12 & Patient & F & 55 & Rural & Bachelor
& Advanced lung cancer
& Inpatient \\

P13 & Patient & M & 56 & Rural & Junior high school
& Advanced lung cancer
& Inpatient \\

P14 & Patient & M & 37 & Urban & Bachelor
& AIDS
& Inpatient \\

P15 & Patient & M & 33 & Rural & High school
& End-stage renal disease, caused by type 2 diabetes
& Home care \\

P16 & Patient & M & 49 & Urban & Illiterate
& End-stage liver cancer
& Inpatient \\

P17 & Patient & F & 71 & Rural & Junior high school
& End-stage chronic heart failure
& Inpatient \\

P18 & Patient & F & 72 & Urban & Junior high school
& Amyotrophic lateral sclerosis (ALS)
& Inpatient \\

P19 & Patient & F & 56 & Rural & High school
& AIDS
& Inpatient \\

P20 & Patient & F & 55 & Urban & Bachelor
& Amyotrophic lateral sclerosis (ALS)
& Inpatient \\

P21 & Patient & M & 44 & Urban & Primary school
& End-stage renal disease, caused by type 2 diabetes
& Home care \\

\bottomrule
\end{tabular}

\end{center}

\section*{Appendix B: Interview Protocols}

\footnotesize
\renewcommand{\arraystretch}{1.18}
\begin{longtable}{p{0.45\textwidth} p{0.50\textwidth}}
\caption{Patient Interview Protocol}\\
\toprule
\textbf{Core Question} & \textbf{Probing Questions} \\
\midrule
\endfirsthead
\toprule
\textbf{Core Question} & \textbf{Probing Questions} \\
\midrule
\endhead
What kinds of health-related digital devices or applications do you currently use in your daily life or care process? & For example, mobile apps, monitoring devices in hospital wards, patient portal websites, or companionship platforms designed for end-of-life care? Were these systems chosen by yourself, or installed and configured by doctors or family members? \\
How often do you usually check the data or information within these systems? & Do you check them yourself, or are they mainly viewed by family members on your behalf? Are there any systems that you almost never actively look at? \\
Under what circumstances do these data usually appear in front of you? & Do you actively open the systems to view the data, or are they automatically displayed or communicated through family members or doctors? Is there any difference between being in hospital and being at home? \\
How were these systems initially installed or activated for you? Did anyone explain what kinds of content would be recorded or who the data would be shared with? & Were you asked for your opinion at the time? Did you have opportunities to ask questions or refuse participation? \\
Among the digital systems you use, are there any types of data or records that you consider private or would prefer others not to see? & What makes those contents feel private to you? Do you perceive different levels of sensitivity among physiological monitoring data, doctor-patient communication records, psychological counseling content, or AI companionship conversations? \\
Do you know who can access the data generated by these systems? & Beyond your doctors and immediate family members, are there other organizations or individuals that might access the data, such as cloud service providers, device manufacturers, or insurance companies? How did you learn this, or are you completely unsure? \\
Your bodily condition is continuously recorded and transmitted through devices. How do you feel about this? & Have there been moments when the way you were ``seen'' through data made you uncomfortable? For example, when family members saw certain data and reacted in ways you did not want? \\
Some apps or platforms automatically generate judgments or assessments based on the information you provide, such as health risk scores or emotional state evaluations. Are you aware of this? What do you think about it? & Are you concerned about where such inferred results might be sent, such as to insurance companies, other medical institutions, or family members? Do you know whether these inferential functions can be reviewed or disabled? \\
Have you ever thought about how your data stored in these systems might be handled after your death? & How would you like these data to be treated, such as deleted, encrypted and archived, or accessible to specific individuals? Are there any data you especially do not want others to see? Have any platforms ever explained this issue to you? \\
Are the accounts and passwords for these systems currently managed by yourself, or have they already been handed over to family members? & When and why did this transfer happen? Was it your active decision, or did you feel there were no other options? After giving up control, did you feel that some content that originally belonged to you became inaccessible? \\
Have you ever tried to adjust privacy settings in these systems, such as restricting access permissions or disabling certain forms of data sharing? & Was the process easy to operate? Did you encounter situations where you could not understand the options or were unsure what would happen after clicking something? Were you ultimately successful? If you gave up, why? \\
If possible, would you prefer different people to access different ranges of your data? For example, allowing doctors to access all health data while limiting some family members to only partial information, or preventing specific individuals from viewing certain records entirely? & Are you currently able to achieve such differentiation? If not, how do you cope with this situation? \\
Have there been any privacy-related concerns that you originally wanted to address but eventually gave up on? & Was it because the process was too complicated, too exhausting, or because you felt it would not make any practical difference? Under your current physical condition, what does this sense of abandonment feel like? \\
If these systems could be redesigned, who would you want to access which kinds of information? Are there any data you would want to remain completely private? & Do you think the idea of ``fully private data'' is realistically achievable? \\
Do you think these systems could be improved in terms of protecting your privacy? & For example, through more fine-grained permission controls, clearer explanations, or by avoiding the recording of certain forms of data altogether? \\
Is there anything else about your experiences with digital systems or about privacy that you would like to add, or that you think is important but has not been discussed? & Open-ended listening without guidance. \\
\bottomrule
\end{longtable}

\begin{longtable}{p{0.45\textwidth} p{0.50\textwidth}}
\caption{Family Member Interview Protocol}\\
\toprule
\textbf{Core Question} & \textbf{Probing Questions} \\
\midrule
\endfirsthead
\toprule
\textbf{Core Question} & \textbf{Probing Questions} \\
\midrule
\endhead
During the process of caring for your family member, what kinds of health-related digital systems or data do you access? & Do you actively check them, or are the data automatically pushed to you by the systems? Were you invited by the patient to participate, encouraged by doctors, or did you intervene because you personally believed it was necessary? \\
Do you currently possess the patient's account passwords or directly manage some systems on their behalf? & When did this begin? How was the decision made? Did the patient explicitly consent, or did it gradually occur without formal discussion? \\
What kinds of data do you usually check, and for what purposes? & Have there been moments when you realized that certain content ``might not have been intended for you to see''? How did you respond in those situations? Did you continue looking? \\
Have there been situations where you felt it was necessary to access or intervene in the patient's digital accounts? & For example, abnormal health indicators, sudden deterioration of the patient's condition, or assisting with medical appointments? Outside such situations, do you also routinely monitor the systems? \\
Do you see viewing the patient's health data as part of caregiving, or as something that also involves the patient's privacy? Have you ever experienced tension between these two aspects? & In your view, how should caregiving needs and the patient's privacy rights be balanced? Are there any forms of data that you believe should remain inaccessible even for caregiving purposes? \\
Have there been moments when you were unsure whether accessing certain data crossed a boundary? & How did you make that judgment? Did you later feel uneasy, or did such situations create friction or awkwardness between you and the patient? \\
Do you know what kinds of information your family member wishes to keep private? & Have you directly discussed this issue together? If not, how do you perceive their boundaries, through their behaviors and expressions, or through your own assumptions? \\
Before viewing or managing the patient's data, do you usually seek their consent? & Did this change after the patient's cognitive or physical capacities declined? When patients cannot clearly express their wishes, who do you think should determine the boundaries of data access? \\
Have you ever encountered situations where you originally intended to check caregiving-related information, but unintentionally accessed much more content than planned? & What motivated you to continue looking? Looking back, how do you evaluate your own behavior at that time? \\
Do the systems you use provide functions that allow patients to determine who can access different kinds of information? & Do these functions meet practical needs? Have you encountered situations where the systems only offered ``fully open'' or ``fully closed'' options without intermediate possibilities? Did this affect your relationship with the patient? \\
Have hospitals, social workers, or doctors ever provided you with explicit guidance regarding the management of patients' digital data? & If not, what do you currently rely on to judge whether your behaviors are appropriate? Would clearer institutional guidance be helpful to you? \\
Have you considered how the patient's digital data should be handled after their death? & Who do you think should decide this? Have any institutions or organizations ever provided guidance on this issue? \\
Is there anything important regarding your experiences with patients' digital data during caregiving that has not been discussed, or that you would like to add? & Open-ended listening. \\
\bottomrule
\end{longtable}

\begin{longtable}{p{0.45\textwidth} p{0.50\textwidth}}
\caption{Physician Interview Protocol}\\
\toprule
\textbf{Core Question} & \textbf{Probing Questions} \\
\midrule
\endfirsthead
\toprule
\textbf{Core Question} & \textbf{Probing Questions} \\
\midrule
\endhead
In your daily clinical practice, what kinds of digital health data related to end-of-life patients do you access? & Beyond electronic medical records within hospital systems, do you also review data generated through third-party apps, wearable devices, or end-of-life care platforms, such as emotional diaries, pain records, or AI companionship conversations? Are these data actively provided by patients, or obtained through other channels? \\
When patients' cognitive or self-care capacities decline, do you recommend or guide family members to become more deeply involved in managing patients' digital accounts? & How is such guidance usually communicated to family members? Are patients simultaneously informed? Is patient consent treated as a necessary condition during this process? \\
Do data generated through consumer health apps or end-of-life care platforms influence your actual clinical decisions? & When using such data, do you consider who the patients originally expected as the audience for those records? For example, when patients complete emotional questionnaires or psychological counseling records, did they anticipate that medical teams would access them? \\
During ward rounds or case discussions, are patients' personal data from third-party platforms directly accessed or discussed? & Before using such data, does the team confirm whether the patient is aware of and consents to their use in clinical contexts? \\
In end-of-life care contexts, how do you think patients' digital privacy rights should be balanced against clinical caregiving needs? & Are there any types of data that you believe should remain private even for medical purposes? Have you encountered situations where privacy and caregiving clearly conflicted, and how were they handled? \\
Some digital platforms automatically generate inferential conclusions based on patients' inputs, such as mortality risk scores, emotional state evaluations, or assessments of ``acceptance of death.'' What are your views on the production and circulation of such data? & Do you think patients should be informed about the existence of these inferential conclusions? Has the possibility of these data flowing to insurance companies or third parties ever been discussed clinically? As physicians, do you believe you have responsibilities to explain such issues to patients? \\
Continuous physiological monitoring enables real-time awareness of patients' conditions. Where do you think the boundaries of such monitoring should lie? & Are there any forms of monitoring that you believe may impose additional psychological or privacy burdens on patients beyond clinical necessity? Have you ever explained the storage and circulation of monitoring data to patients? \\
Are you aware of how patients' digital data are handled after death across various platforms? & Do you believe current laws, regulations, or institutional policies adequately address these issues? Have patients or family members ever raised concerns about this, and how did you respond? \\
Does your medical institution have explicit protocols or operational guidelines regarding access to and management of patients' digital data? & If not, what do you rely on in practice to judge whether certain forms of data access are appropriate? When encountering ambiguous situations, do you tend to maintain the status quo or seek further confirmation? \\
Has your department or hospital ever discussed establishing dedicated policies or agreements regarding access to patients' digital data? & If such discussions occurred, why were they not implemented? Was it due to legal complexity, clinical workload, or lack of institutional priority? What do you personally consider the major barriers to advancing such governance? \\
Have you ever received professional training related to protecting patients' digital privacy? & Do you think there are significant deficiencies in current medical education or continuing education regarding these issues? If improvements were made, what aspects would be most important? \\
In situations where patients, family members, and medical teams jointly participate in digital data management, have you observed conflicts or problems caused by unclear permission boundaries or information access? & Could you provide a specific example? How was the issue handled? In your opinion, are the root causes primarily technical, institutional, or communicative? \\
Is there anything important regarding digital privacy in end-of-life care that has not been covered, or any observations or perspectives you would like to add? & Open-ended listening. \\
\bottomrule
\end{longtable}

\clearpage
\twocolumn

\section*{Appendix C: Meta-Review}

The following meta-review was prepared by the program committee for the IEEE Symposium on Security and Privacy (S\&P) as part of the review process.

\subsection*{Summary of Paper}

This paper studies end-of-life digital privacy. The study uses multi-stakeholder semi-structured interviews (involving terminally ill patients and their caregivers and doctors), to identify times where digital privacy is relevant to terminally ill patients and patients, caregiver, and doctor attitudes toward privacy in these contexts.

\subsection*{Scientific Contributions}

\begin{itemize}
    \item Establishes a New Research Direction
    \item Provides a Valuable Step Forward in an Established Field
\end{itemize}

\subsection*{Reasons for Acceptance}

\begin{enumerate}
    \item This paper addresses an important aspect of privacy for an under-studied population. While multiple stakeholders are considered, the study is centered around the terminally ill patients, their data, and their care.

    \item The study is well-designed and the methods are thoroughly and clearly described. The interview questions are aligned with the goals of the study and the analysis was performed appropriately, including explaining in detail how data saturation was determined.

    \item This paper addresses a sensitive subject with a vulnerable population. The authors include a thorough discussion of ethical considerations, including steps taken to minimize the risks to participants as well as the study team.
\end{enumerate}
\normalsize

%% file: sections/appendix_demographics.tex
\normalsize